\documentclass[prd,preprint,showpacs,showkeys,preprintnumbers]{revtex4}
\usepackage{graphicx}                                
\usepackage{amsmath,amsfonts,bbm}
\graphicspath{{./Figures/}}                          

\usepackage{epsfig}
\usepackage{amssymb}
\usepackage{amsfonts}
\usepackage{float}
\newcommand{\Tr}{\mbox{Tr}}

\newcommand{\beq}{\begin{equation}}
\newcommand{\eeq}{\end{equation}}
\newcommand{\bea}{\begin{eqnarray}}
\newcommand{\eea}{\end{eqnarray}}
\newcommand{\beas}{\begin{eqnarray*}}
\newcommand{\eeas}{\end{eqnarray*}}
\newcommand{\eq}{\begin{equation}}
\newcommand{\en}{\end{equation}}
\newcommand{\eqa}{\begin{eqnarray}}
\newcommand{\ena}{\end{eqnarray}}

\begin{document}

\preprint{}

\title{Coulomb gauge gluon propagator and the Gribov formula}
\author{G.~Burgio}
\author{M.~Quandt}
\author{H.~Reinhardt}
\affiliation{Institut f\"ur Theoretische Physik, T\"ubingen, Germany}

\begin{abstract}
We analyze the lattice SU(2) Yang-Mills theory in Coulomb gauge. We show that
the static gluon propagator is multiplicative renormalizable and takes the 
simple form $D(|\vec{p}|)^{-1}=\sqrt{|\vec{p}|^2+M^4/|\vec{p}|^2}$, proposed 
by Gribov through heuristic arguments many years ago. 
We find $M=0.88(1) {\rm GeV} \simeq 2 \sqrt{\sigma}$.
\end{abstract}

\maketitle

\section{Introduction}
\label{sec:introduction}

Non-Abelian gauge theories in Coulomb gauge have 
recently been the subject of extensive research. In particular, estimates for 
the ground state wave function have been derived through variational methods
\cite{Szczepaniak:2001rg,Feuchter:2004mk}. A key ingredient
in such procedure is the static transverse gluon propagator 
$D(|\vec{p}|)$, which turns out to be compatible 
with the Gribov-Zwanziger scenario \cite{Gribov:1977wm,Zwanziger:1995cv}. 
Unfortunately up to now {\it ab initio}
lattice calculations have failed to deliver a conclusive 
comparable result. A first study carried out for SU(2) at fixed $\beta=2.2$
\cite{Cucchieri:2000gu,Cucchieri:2000kw} indicated compatibility with 
Gribov's formula 
\cite{Gribov:1977wm} in the IR but was inconclusive 
in the UV. Later studies in SU(2) and SU(3) showed 
strong scaling violations and contradictory IR results for $D(|\vec{p}|)$
\cite{Langfeld:2004qs,Quandt:2007qd,Voigt:2007wd}. All featured a UV 
behaviour at odds with 
simple dimensional arguments. Except for \cite{Cucchieri:2000gu,%
Cucchieri:2000kw}, where 
the pole structure of the full propagator was analyzed,
all these studies \cite{Langfeld:2004qs,Quandt:2007qd,Voigt:2007wd} have 
simply calculated the static propagator relying on the fact that 
once Coulomb gauge is fixed $D(|\vec{p}|)= 
\sum_{p_0 }D(|\vec{p}|,p_0)$ does not depend on the residual gauge freedom 
of temporal fields. One takes, 
however, {\it de facto} for granted that multiplicative renormalizability
holds for the full propagator, although perturbative results seem to 
point at a more complex picture \cite{Watson:2007mz}, and that cutoff effects 
do not influence the result. To verify these assumptions 
we will fix analytically the residual temporal gauge
and study the renormalization of the full propagator $D(|\vec{p}|,p_0)$. 
We will show that: i) the latter is not multiplicatively 
renormalizable; 
ii) the static propagator $D(|\vec{p}|)$ is renormalizable only in the limit 
of continuous time, i.e. in the lattice Hamiltonian formulation. The resulting 
$D(|\vec{p}|) = \int_{-\infty}^{+\infty}d p_0\,D(|\vec{p}|,p_0)$ satisfies the
Gribov formula for all momenta \cite{Gribov:1977wm}.

\section{Setup}
\label{sec:setup}

We discretize the SU(2) Yang-Mills theory through the 
Wilson plaquette 
action $S=\beta \sum_{x,\mu\nu} (1-\frac{1}{2}\,\Tr\, U_{\mu\nu}(x))$, 
employing the standard
Creutz algorithm (heath-bath plus over-relaxation). We take $2.15\leq\beta\leq
2.6$ and lattices $L$=12, 16, 20 and 24, with 200 to 1000 
measurements, depending on $L$.
The lattice gluon propagator $~D^{ab}_{ij}(p)~$ is the
Fourier transform of the gluon two-point function:
\beq
D^{ab}_{ij}(p) =
\left\langle \widetilde{A}^a_{i}({k}) \widetilde{A}^b_{j}(-{k})
\right\rangle_U
   = \delta^{ab} \delta_{ij} D(p)\;, \qquad  
p_{\mu} = p({k}_{\mu}) = \frac{2}{a} \sin\left(\frac{\pi
      {k}_{\mu}}{L}\right)\;.
\label{eq:D-def}
\eeq
Here $\widetilde{A}^a_{\mu}({k})$ is the Fourier transform of the lattice 
gauge
potential $A^a_\mu(x+\hat{\mu}/2)$ (see Eq.~(\ref{eq:transversality})), 
$a$ is the lattice spacing, $p=(\vec{p},p_0)$ denotes the four-momentum and
$~{k}_{\mu} \in (-L/2, +L/2]$ are the integer-valued lattice momenta.
Results for the temporal component $D^{ab}_{00}(p)$ will be presented in a 
forthcoming paper \cite{prep}.
The Coulomb gauge is fixed for each configuration maximizing for every time
slice
\beq 
F_g(t) = \frac{1}{6 L^3} \sum_{\vec{x},i} \mbox{Tr} ~
U^g_{i}(\vec{x},t)\;,\qquad
U^g_{i}(\vec{x},t) = g(\vec{x}) ~U_{i}(\vec{x},t) 
~g^{\dagger}(\vec{x}+{\hat{i}})
\label{eq:functional}
\eeq
with respect to local gauge transformations $g(\vec{x}) \in$ SU(2). The local 
maxima of $~F_g(t)~$ satisfy for each $t$ the differential 
lattice Coulomb gauge transversality condition for the gauge potentials:
\beq
\partial_{i} A^g_{i}(\vec{x},t) =
       A^g_{i}(\vec{x}+{\hat i}/2,t)-A^g_{i}(\vec{x}-{\hat{i}}/2,t) = 0\;,\quad
A_{\mu}(x+ {\hat \mu}/2) =
       \frac{1}{2i}(U_{\mu}(x)-U_{\mu}(x)^{\dagger})\;.
\label{eq:transversality}
\eeq
In detail we: i) apply for the whole lattice a flip preconditioning 
\cite{Bogolubsky:2005wf,Bogolubsky:2007bw} as in Ref.~\cite{Quandt:2007qd};
ii) maximize $F_g(t)$ via several standard over-relaxation and/or simulated 
annealing cycles after a random gauge copy for each time separately; 
iii) choose at every slice the copy with the highest $F_g(t)$; iv) perform 
a random gauge transformation on the whole lattice and repeat from step i). 
The number of total recursions varies between 10-40, depending on $\beta$ and 
$L$. The copy of the whole lattice with the best functional value 
$\sum_t F_g(t)$ is chosen to select the gauge-fixed configuration. This 
procedure remains relatively efficient for every volume at higher $\beta$, 
but in the strong coupling (low $\beta$) region Gribov noise is no longer 
under control as the lattice size increases. This prevents us from going 
beyond $L=24$. Future studies on larger lattices should rather adapt
the improved algorithm proposed in Ref.~\cite{Bogolubsky:2007bw}. 

Once the Coulomb gauge has been fixed, only space independent 
gauge transformations $g(t)$ can be performed at each time slice. 
In the continuum and on the lattice one can choose:
\beq
0=\int\, d^3x\, \partial_\mu A_\mu(\vec{x},t) \Rightarrow \partial_0 \int\, d^3x\, A_0(\vec{x},t) = 0 \quad \Leftrightarrow\quad u(t)\,=\frac{1}{L^3}
\sum_{\vec{x}} U_0(\vec{x},t) \to {\rm const.}
\label{eq:6}
\eeq 
Define $\hat{u}(t) = {u(t)}/{{\mbox{Det}}\left[u(t)\right]} \in$ SU(2).
Periodic boundary conditions  make
$\Tr \prod_t \hat{u}(t) = \Tr\,P$ invariant under $g(t)$.
Eq.~(\ref{eq:6}) can thus be fixed recursively through
$g(t) \hat{u}(t) g^{\dagger}(t+1) = \tilde{u}\equiv P^{{1}/{L}}$,
up to a $g(0)$ satisfying $\left[g(0),P\right]=0$. We can choose 
$g(0)=\mathbbm{1}$. 
Of course $\tilde{u} \to \mathbbm{1}$ only for $L \to \infty$. 
We find $1-{\mbox{Tr}}\, \tilde{u}\simeq 10^{-3}$ already for $L=24$, with 
leading corrections $\propto L^{-2}$. 

We always select 
spatial momenta indices
$\vec{k}$ satisfying a cylinder cut \cite{Leinweber:1998uu,Voigt:2007wd} to 
minimize 
violations of rotational invariance. Time-like momenta are unconstrained.
The scale is set re-expressing the lattice spacing $a(\beta)$ in units of the 
string tension $\sigma=(440 {\rm MeV})^2$ \cite{Bloch:2003sk}.

\section{Results}
\label{sec:results}

\begin{figure}
\includegraphics[width=1.0\textwidth,height=1.0\textwidth]{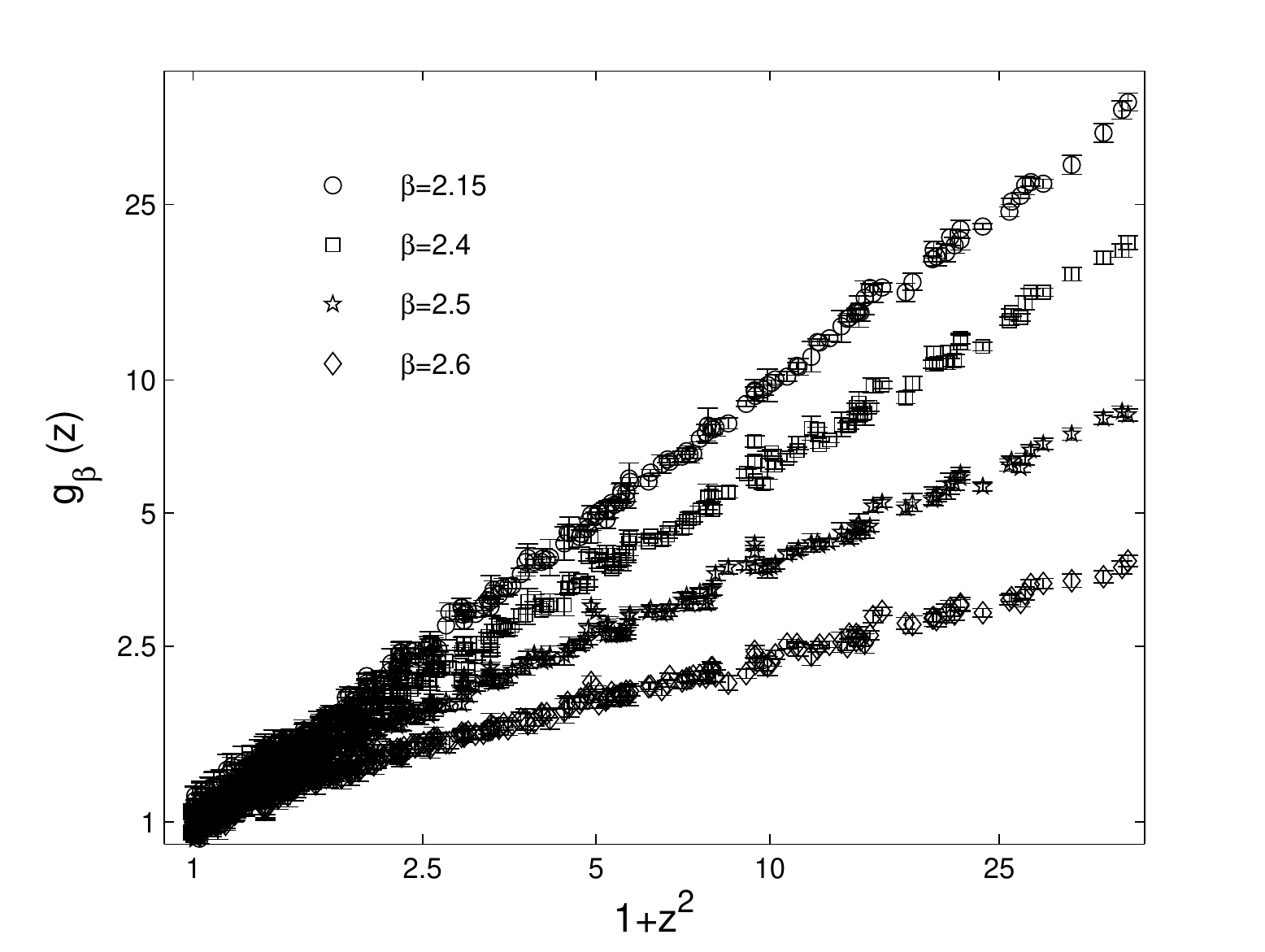}
\caption{Data for $g_\beta(z)$ vs $1+z^2$ in log-log scale, $L=24$. 
For sake of readability not all $\beta$ are shown.}
\label{fig:polaw}
\end{figure}

We find that the lattice bare propagator 
$D_\beta(|\vec{p}|,p_0)$ satisfies over the whole $p$ range:
\beq
D_\beta(|\vec{p}|,p_0) = \frac{f_\beta(|\vec{p}|)}{|\vec{p}|^2}
\frac{{g}_\beta(z)}{1+z^2}\qquad z=\frac{p_0}{|\vec{p}|}\;,
\label{eq:fact}
\eeq
where $|\vec{p}|^2(1+z^2)$ is explicitly factorized since for large $p^2$ one 
expects $D_\beta(|\vec{p}|,p_0) \simeq (|\vec{p}|^2+p_0^2)^{-1}$ on 
dimensional grounds. $f_\beta$ and $g_\beta$ are dimensionless functions and
without loss of generality we can choose $g_\beta(0)= 1$. From 
Eq.~(\ref{eq:fact}) it is clear that a dynamical mass generation must occur
in $f_\beta$, i.e. it must be a function of $|\vec{p}|/M$.
We can check that taking $\sum_{p_0} D_\beta(|\vec{p}|,p_0)$ 
reproduces the results of Ref.~\cite{Quandt:2007qd}. The data for 
${g}_\beta(z)=(1+z^2)D_\beta(|\vec{p}|,p_0)/D_\beta(|\vec{p}|,0)$ 
are shown in Fig.~\ref{fig:polaw} for $L=24$. Their leading behaviour
is well described by a power law $(1+z^2)^{\alpha}$. 
Sub-leading 
corrections can be parameterized equally well through different Ans\"atze,
with all fits giving $\chi^2$/d.o.f. between 0.8 and 1.2. Without further 
theoretical input we have no reason to prefer one upon the other.
For $\beta \gtrsim 2.3$ the functions $g_\beta$ vary quite consistently with 
$L$. 
In Table~\ref{tab:polaw} we give our best estimates for $\alpha(\beta)$ in 
the large $L$ limit, assuming again a leading correction $\propto L^{-2}$. 
Within one or two standard deviations all values of $\alpha$ are compatible 
with 1, i.e. 
$D_\beta(|\vec{p}|,p_0)$ might be $p_0$ independent in the
thermodynamic limit. A study on larger,
perhaps anisotropic lattices will be needed to clarify the matter.
\begin{table*}[htb]
\begin{center}
\mbox{
\begin{tabular}{|r|c|c|c|c|c|c|c|}
\hline
$\beta$ & 2.15 & 2.2 & 2.25 & 2.3 & 2.4 & 2.5 & 2.6 \\
\hline\hline
$\alpha(\beta)$& 1.01(2) & 1.02(3) & 1.02(3) & 0.99(5) & 0.96(6) & 0.8(1) &0.6(2)\\
\hline
$Z(\beta,\mu)$&0.403(4)&0.430(5)&0.456(6)&0.475(7)&0.515(10) &0.536(12)&0.548(13)\\
\hline
\end{tabular}
}
\end{center}
\caption{$L\to\infty$ exponents $\alpha$ for the functions $g_\beta(z)$; 
renormalization coefficients $Z(\beta,\mu)$ as in Eq.~(\ref{eq:ren}).}
\label{tab:polaw}
\end{table*}
It is anyway clear from Eq.~(\ref{eq:fact}) that, as long as for our choices 
of $\beta$ and $L$ the functions $g_\beta(z)$ 
remain different, as they do, $D_\beta(|\vec{p}|,p_0)$ {\it can not} be 
multiplicative 
renormalizable. On the other hand, let us consider different lattice 
cut-offs for space and time $a_s/a_t =\xi>1$ \cite{Burgio:2003in}. 
For $\xi \to \infty$ (lattice Hamiltonian 
limit) $D_{\beta}(|\vec{p}|) \propto |\vec{p}|^{-1} {f}_\beta(|\vec{p}|)$ 
up to the (possibly diverging) constant $\int_0^\infty d z
g_\beta(z)/(1+z^2)$ and multiplicative 
renormalizability relies solely on $f_\beta(|\vec{p}|)$.
Eq.~(\ref{eq:fact}) also explains the scaling violations in 
Ref.~\cite{Langfeld:2004qs,Quandt:2007qd,Voigt:2007wd}. 
Let us assume that $f_\beta(|\vec{p}|)$ is renormalizable, define 
$\hat{p} = a_s |\vec{p}|$ and for simplicity neglect
sub-leading corrections in $g_\beta(z)$. For 
large $L$ we can approximate the discrete sum over $p_0$
by an integral and:
\bea
D_{\beta}(|\vec{p}|)&\simeq& \int_{-\frac{2}{a_t}}^{\frac{2}{a_t}} 
\frac{d p_0}{2 \pi}D_\beta(|\vec{p}|,p_0)=\frac{f_\beta(|\vec{p}|)}{|\vec{p}|}
\int_{0}^{\frac{2 \xi}{\hat{p}}} \frac{d z}{\pi} 
\left(1+z^2\right)^{\alpha-1} = \frac{f_\beta(|\vec{p}|)}{|\vec{p}|}
I(\frac{2 \xi}{\hat{p}},\alpha)\;;\nonumber \\
I(\frac{2 \xi}{\hat{p}},\alpha) &=& \frac{1}{2 \pi}
B(\frac{4 \xi^2}{4 \xi^2+\hat{p}^2},\frac{1}{2},-\alpha+\frac{1}{2})\;,
\label{eq:ff}
\eea
where $B(z,a,b)$ is the incomplete beta function. 
For $\xi\to\infty$ $I$ explicitly goes to a $|\vec{p}|$ 
independent constant, $B(1,1/2,-\alpha+1/2)$, while in a standard lattice 
formulation, where $\xi\equiv 1$, 
there is no way to avoid the extra $|\vec{p}|$ dependence
$I({2}/{\hat{p}},\alpha)$, which will 
violate multiplicative renormalizability. Indeed such
function times the tree level result $|\vec{p}|^{-1}$ 
compares well with the behaviour observed in
\cite{Langfeld:2004qs,Quandt:2007qd,Voigt:2007wd}.

\begin{figure}
\includegraphics[width=1.0\textwidth,height=1.0\textwidth]{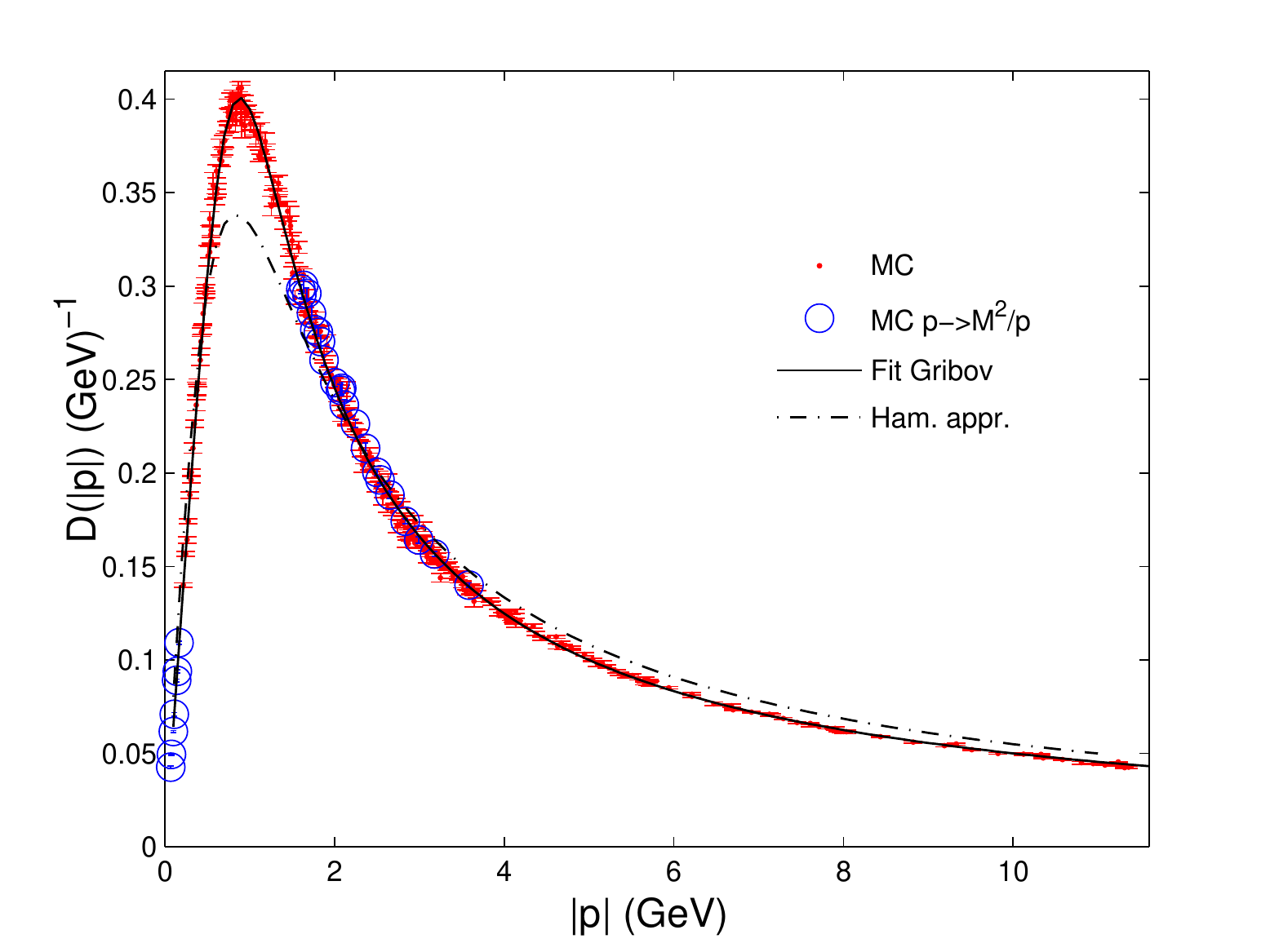}
\caption{
Gluon propagator. Circles are (a few) data for 
$|\vec{p}|\to M^2/|\vec{p}|$, the continuum line is the fit to Gribov's 
formula and
the dot-dashed line the result of the Hamiltonian approach of 
Ref.~\cite{Epple:2006hv}.}
\label{fig:glprp}
\end{figure}

From the above discussion it is clear that the $p_0$ dependence
is immaterial to the static propagator in the Hamiltonian limit,
all the relevant information being encoded in $f_\beta(|\vec{p}|)$. Since
the limit $\xi\to\infty$ is out of the reach of available simulations, to 
extract $f_\beta$ from simulations at standard sizes and coupling 
(i.e. spacing) one could use only results at fixed $p_0$, 
e.g. $f_\beta(|\vec{p}|)\equiv D_\beta(|\vec{p}|,0)$, at the cost of 
loosing data. Although this works, we propose here an alternative 
procedure which makes use of the results for all energies, increasing the 
statistics.
We divide the $z$ dependence out of the data using the fitted functional form 
for $g_\beta(z)$ and average the result over $p_0$:
\beq
f_\beta(|\vec{p}|) := 
\frac{1}{L}\sum_{p_0} D_\beta(|\vec{p}|,p_0)\frac{1+z^2}{{g}_\beta(z)}\,;
\quad
D_{\beta}(|\vec{p}|):=\frac{{f}_\beta(|\vec{p}|)}{|\vec{p}|}\;.
\label{eq:enind}
\eeq 
The factor $|\vec{p}|^{-1}$ only serves to restores dimensions when comparing 
with the Hamiltonian result, according to Eq.~(\ref{eq:ff}), and is immaterial
to renormalization. 
To minimize finite size effects we exclude spatial momenta not
satisfying a cone cut \cite{Leinweber:1998uu,Voigt:2007wd} for all lattices 
with $(a(\beta) L)^4 < (2.5 {\rm fm})^4$. 
The $D_{\beta}(|\vec{p}|)$ calculated form Eq.~(\ref{eq:enind}) are 
multiplicative renormalizable via 
$D_{\mu}(|\vec{p}|)|=Z(\beta,\mu)D_{\beta}(|\vec{p}|)$, where $\mu$ defines 
the renormalization point. We choose it 
such that (see Table~\ref{tab:polaw}):
\beq
D_{\mu}(|\vec{p}|)|_{\mu=\infty}=Z(\beta,\mu)|_{\mu=\infty}D_{\beta}
(|\vec{p}|)|\underset{|\vec{p}| \to \infty}{\to}\frac{1}{2 |\vec{p}|}\;,
\label{eq:ren}
\eeq
corresponding to the naive UV continuum behaviour. 
Fig.~\ref{fig:glprp} shows 
the result of our procedure. 
We have performed 
fits of $D_{\mu}(|\vec{p}|)$ allowing at the same time
a power law in the IR, $|\vec{p}|^{q}/M^{2-q}$, and a power 
law plus logarithmic corrections in the UV, 
$|\vec{p}|^{-r}|\log{|\vec{p}|}^{-s}$. We
get $q = 0.99(1)$, $r = 1.002(3)$, $s=0.002(2)$ and $M=0.88(1)$ GeV, 
with $\chi^2$/d.o.f. all in the range 3.2-3.3, in agreement with the
UV and IR analysis in 
Ref.~\cite{Feuchter:2004mk,Epple:2006hv,Schleifenbaum:2006bq}. Since 
our data show excellent IR-UV symmetry for $p\leftrightarrow M^2/p$ (see 
Fig.~\ref{fig:polaw}), we constrain 
$q=r=1$, $s=0$ and fit the whole result through the Gribov formula
\beq
D_{\mu}(|\vec{p}|)|_{\mu=\infty}=\frac{1}{2 \sqrt{|\vec{p}|^2+
\frac{M^4}{|\vec{p}|^2}}}
\eeq
We find just as good agreement ($\chi^2$/d.o.f. $= 3.3$)
again with $M=0.88(1) {\rm GeV} \simeq 2 \sqrt{\sigma}$. This
compares well with $M\simeq 0.5-0.8$ GeV from the IR analysis in 
Ref.~\cite{Cucchieri:2000gu,Cucchieri:2000kw}, 
given that only $\beta=2.2$ was used there. 
Apart from 
unavoidable residual violations of rotational invariance, the highest noise 
in the data occurs around the peak, $|\vec{p}|\simeq M$. 
Only better gauge fixing 
and larger volumes can help here. We have also 
collected data for $2.6 < \beta \leq 2.8$. They are still in perfect 
agreement with the above formula, however since physical volumes are 
very small we cannot claim to have control over systematic effects, so 
we omit them in the above fits. An investigation on higher volumes would
be welcome.

\section{Conclusions}
\label{sec:conclusions}

We have shown that on the lattice 
the static gluon propagator in Coulomb gauge is multiplicative 
renormalizable only in the Hamiltonian limit. 
This calls for analogous investigations in the
continuum, where to our knowledge renormalizability has not been 
proven in Coulomb gauge.
We have also given a procedure to extract the relevant information from 
the data at accessible lattice sizes and spacings. Summarizing this procedure: 
i) calculate the full bare propagator $D_\beta(|\vec{p}|,p_0)$ after fixing 
the residual 
gauge as in Eq.(\ref{eq:6}); ii) fit to the data the function $g_\beta(z)$ 
arising from the factorization in Eq.(\ref{eq:fact}); iii) divide out 
$g_\beta(z)$ from $D_\beta(|\vec{p}|,p_0)$ to isolate its $p_0$ independent
part $f_\beta(|\vec{p}|)$; iv) define $D_\beta(|\vec{p}|)=
f_\beta(|\vec{p}|)/|\vec{p}|$.
The so obtained static propagator $D_\beta(|\vec{p}|)$ is multiplicative 
renormalizable, UV-IR 
symmetric and is well described by Gribov's formula over the whole 
momentum range. Its infrared and ultraviolet behaviours are in good
agreement with the results obtained in the variational approach to 
continuum Yang-Mills theory in Coulomb gauge \cite{Feuchter:2004mk}. 
Both the infrared analysis of the Dyson-Schwinger equations in this 
approach and its numerical solution yield for the 
infrared exponent $q = 1$ \cite{Epple:2006hv,Schleifenbaum:2006bq}. 
We will compare further quantities obtained in the
variational approach \cite{Feuchter:2004mk} with the lattice
using the procedure proposed above \cite{prep}. 

\section*{ACKNOWLEDGMENTS}

We would like to thank Peter Watson for 
stimulating discussions. This work was partly supported by DFG under 
contracts DFG-Re856/6-1 and DFG-Re856/6-2.
\bibliographystyle{apsrev}
\bibliography{references}

\end{document}